\documentclass{article}

\usepackage{PRIMEarxiv}

\usepackage[utf8]{inputenc} % allow utf-8 input
\usepackage[T1]{fontenc}    % use 8-bit T1 fonts
\usepackage{hyperref}       % hyperlinks
\usepackage{url}            % simple URL typesetting
\usepackage{booktabs}       % professional-quality tables
\usepackage{amsfonts}       % blackboard math symbols
\usepackage{nicefrac}       % compact symbols for 1/2, etc.
\usepackage{microtype}      % microtypography
\usepackage{fancyhdr}       % header
\usepackage{graphicx}       % graphics
\graphicspath{{media/}}     % organize your images and other figures under media/ folder

\title{Vulcan and anomalous displacement of Mercury's perihelion}

\author{
	Pogossian S.P.\\
	Univ. Brest, CNRS, IRD, Ifremer, IUEM, Laboratoire d'Oc\'eanographie Physique et Spatiale (LOPS),\\ Technopôle Brest-Iroise, Rue Dumont d'Urville, 29280 Plouzan\'e \\
	Universit\'e de Bretagne Occidentale \\
	Brest, France\\
	\texttt{pogossia@univ-brest.fr} \\
}

\begin{document}
	\maketitle

	\begin{abstract}
		\ In this paper, I re-examine the question of a possible explanation of the anomalous advance of Mercury's perihelion by the existence of the hypothetical planet Vulcan proposed by Le Verrier, whose orbit would be located inside the orbit of Mercury. My calculations are focused on the optimization of the orbital parameters of Vulcan in order to explain precisely the anomalous advance of Mercury's perihelion. To reach this goal, I used recent experimental results concerning the observations of the intra-mercurian zone. My calculations establish the direct relation of the anomalous advance of Mercury's perihelion with the mass of Vulcan and its distance to Mercury.
	\end{abstract}

	% keywords can be removed
	\keywords{Asteroid belt \and General gravity \and Ephemerides \and Planets and satellites: individual (Mercury) \and (Stars): planetary systems.}

	\section{Introduction}
	\
	The measured anomalous advance of Mercury's perihelion of $43^{\prime\prime}$/cy led Le Verrier to propose the existence of another hypothetical planet within Mercury's orbit that he named Vulcan \cite {leverrier1860}, \cite {laskar2017}, whose gravitational influence on Mercury would explain the observed anomalous advance. 
	
	Although Mercury is the first planet in the solar system according to Titius Bode's law, due to the modifications made by Jenkins \cite {jenkins1878} to this law, the latter would allow the existence of another planet between Mercury and the Sun with an orbit of major half-axis equal to 0.1 AU.  From this date, an intense research of this planet has been initiated, but with some false discoveries: \cite {levenson2016}, \cite {baum1997}.
	
	Even if the explanation of this phenomenon by Einstein based on general gravitation have significantly slowed down the search for this hypothetical planet, observations are still made to this day in the area likely to host this planet, inside the orbit of Mercury.
	
	In 1978, Weidenschilling  \cite{weidenschilling1978} proposed the possible presence of intra-mercurian planetesimal orbits as a possible explanation for Mercury's low mass and high iron content compared to other telluric planets.  Studying the evolution of Mercury,  Leake et al. \cite{leake1987} put forward the idea that the existence of a population of planetesimals specifically bombarding Mercury inside its orbit and only occasionally bombarding Mercury due to secular perturbations could be one of the important sources of its early cratering.
	
	However, these mobile objects whose orbits may cross that of Mercury will exhaust themselves through mutual collisions in only ~1 Gyr. Campins et al \cite{campins1996} have argued for the possible existence of 1-km diameter vulcanoids in a restricted elliptical range of 0.1 to 0.25 AU from the Sun.
	
	The calculations of Evans and Tabachnik \cite{evans1999} showed that the zone where planetesimals of radius higher than 0.1 km can survive during the age of the solar system, is located between 0.09 and 0.21 AU between the Sun and Mercury.  With however deviations at 0.15 and 0.18 AU corresponding to resonances of average motion with Mercury and Venus \cite{evans2002}. Since the evolution of the vulcanoid population due to close encounters is slower at large heliocentric distances, vulcanoids of size 1 km or larger are more likely to be found in the region between 0.16 and 0.18 AU. 
	
	The study of the population of  vulcanoids by Vokrouhlicky et al \cite{vokrouhlicky2000} by examining the effect of Yarkovsky on the drift of their major half-axis put a limit to their number : 300-900 bodies of more than 1 km in diameter.  
	
	It should be noted that the observations of possible vulcanoids are very difficult to realize because of the solar glare. In order to find vulcanoids on heliocentric orbits of 0.07 to 0.21 AU inside Mercury's orbit, the study of SOHO/LASCO C3 coronagraph images undertaken by Durda et al \cite{durda2000}, did not reveal any moving objects of 20 and 60 km in diameter at the inner and outer limits of the vulcanoid zone. Exploring the effects of collisional evolution of vulcanoids, Stern and Durda  \cite{stern2000} came to the conclusion that the region between 0.06 and 0.21 AU from the Sun could be populated by vulcanoids and that the most favorable location for the survival of mobile bodies is probably near the outer edge of dynamical stability (0.2 AU). According to these authors \cite{stern2000}, there will be no more than a few dozen objects of radius greater than 1 km in a ring up to the stability limit of 0.2 AU.  
	
	By analyzing images taken by the coronograph LASCO on board the satellite SOHO, Schumacher and Gay \cite{schumacher2001} did not find any object with a diameter greater than 60 km. 
	
	During the total solar eclipse, Zhao et al \cite{zhao2009} investigated the possible presence of moving objects in the area extending from about 0.08 AU to 0.18 AU within Mercury's orbit, but no vulcanoids were detected up to a diameter limit of 2-6 km.
	
	The study by Steffle et al.\cite{steffl2013} concerns the search for vulcanoids inside Mercury's orbit, in the dynamically stable zone between 0.07 and 0.21 AU using archival data from the Heliospheric Imager-1 (HI-1) instrument on the two NASA STEREO spacecraft. In the obtained images, no vulcanoids were detected with the orbital parameter $e<0.15$  and $\textit{i} < 15^\circ$. This allowed them to conclude that there are currently no vulcanoids larger than 5.7 km in diameter and that there are no more than 76 vulcanoids larger than 1 km in diameter. 
	
	In a review paper, Beech and Peltier \cite{beech2017} conclude that while there is not enough data to completely rule out the eventual existence of a small population of vulcanoids, observations have drastically reduced their possible survival in the intra-mercurian orbit area. 
	
	Thus, in spite of the considerable efforts of the researchers of the whole world since more than one century, and the important means implemented to seek the planet Vulcan, in particular the research carried out by NASA, nothing was found which would approach the characteristics predicted by Le Verrier, namely a mass of approximately 1/17 of the mass of Mercury, a semimajor axis of the order of \textit{a}=0.14 UA and an inclination of about $12^\circ$.  The impossibility to explain the anomalous advance of Mercury's perihelion by the existence of a single planet Vulcan, leaves open another possibility, that of an intra-mercurian asteroid belt of fairly large total mass. But, the existence of a large number of intra-mercurian asteroids is rendered practically improbable by all the observations made so far.
	
	Thus the planet Vulcan has remained a myth. It is therefore legitimate to wonder if this planet imagined by Le Verrier could really explain the APM anomaly of Mercury without using general relativity. In order to answer this question and on the basis of the set of observations whose goal was the detection of Vulcan, I have optimized the orbital parameters and the mass of the hypothetical Vulcan allowing a perturbation large enough to explain the anomalous APM, i.e. $43^{\prime\prime}/cy$ \cite{pogossian2022}.{\tiny }

	\section{Calculations}
	\label{sec:2}
		
	The Newtonian gravitational equations for a 10-body problem (including the Sun and Vulcan) have been integrated over a time interval of 365\,260 days with a time step of about 60 minutes.  It is noteworthy that by using this integration time interval of about 1\,000 years, we will take into account a possible effect of the quasi commensurability of the periods of Jupiter and Saturn known as the great inequality and which has a period of about 900 years.
	
	To integrate a 10-body solar system (Sun and Vulcan included), one needs to know the values of the basic parameters of the 10-body problem: the planet/sun mass ratio, the Newtonian constant of universal gravitation G, as well as the initial values of the positions and velocities of all the planets and of the sun. For the mass of Vulcan I have chosen values in the following interval: from negligible mass up to a mass of 3.3 mass of Mercury including of course the specific value chosen by Le Verrier, i.e. 1/17 of the mass of Mercury. Its initial heliocentric position and velocity have been determined by an optimization procedure around orbital elements close to the values reported by Le Verrier with the constraints imposed by modern observation data.
	
	Although the most available ephemerides are based on experimentally measured data, they include corrections based on general relativity, since the values of all parameters depend on model-dependent optimization procedures. Our calculations require Newtonian ephemeris data without relativistic corrections. Le Guyader used an optimization program to subtract the relativistic corrections and the influence of the Moon that correspond to particular values of the gravitational constants (GM) taken from the DE200/LE200 ephemerides \cite{leguyader1993}. This type of initial values has also been reported by  Lieske \cite{lieske1967}, by Arminjon \cite{arminjon2002} and \cite{arminjon2004}.
	
	In this study, the initial positions and velocities of the sun and the 8 planets at the date of the Julian ephemeris JJ = 2\,451\,600.5 taken from the reference \cite{leguyader1993} were chosen, as for the gravitational constants and the astronomical unit, their values as reported in the ephemeris DE200/LE200 \cite{leguyader1993}, \cite{standish1990} and which were used by Le Guyader at the time of the subtraction of the relativistic corrections were used.
	
	A MATLAB ODE113 solver was used with RelTol=$3\times10^{-14}$  and  AbsTol=$10^{-16}$, values slightly different from those recommended by Arminjon \cite{arminjon2004} but which provide better accuracy for the initial data first used by Le Guyader \cite{leguyader1993}.  The initial position and the initial velocity of the hypothetical planet Vulcan were calculated from the orbital elements on the basis of modern observations of the intra-mercurian region and on the data suggested by Le Verrier.  The optimization procedure consisted in selecting the orbital elements giving the greatest increase of the advance at Mercury's perihelion.
	
	The integration results were analyzed in an invariant coordinate system related to the initial value of the barycentric linear momentum vector of the 9 planets $\vec{P}_{i}=\sum_{i=2}^{10} M_{i} \vec{v}_{i}$, and the total angular momentum vector $\vec{L}=\sum_{i=1}^{10} \vec{r}_{i}\times\vec{P}_{i}$  of the system of the 10 bodies \cite{souami2012}.  
	
	The Z axis is directed by the direction of the total angular momentum  $\vec{L}$. Two other vectors, $\vec{C}$ and  $\vec{D}$ that are related to $\vec{P}$ and  $\vec{L}$ can now be defined by the following relations: 
	$\vec{C}=\vec{P}\times\vec{L}$  and  $\vec{D}=\vec{C} \times \vec{L}$. It is assumed that the X and Y axes are directed respectively along  $\vec{D}$ and  $\vec{C}$. The barycenter of the solar system, composed of 10 bodies in our calculation, is taken as the origin of our invariant reference system.
	
	The maximum integration error of the MATLAB ODE113 solver has been analyzed previously \cite{arminjon2004}, \cite{leguyader1993}, \cite{roy2014}, by comparing the difference between the initial values of the planetary coordinates and velocities and their values after back and forth  time integration. The integration method for the study of the advance of Mercury's perihelion and the accuracy of the integration was evaluated in a previous work \cite{pogossian2022} that demonstrates the reliability of the calculation method. 
	
	In the work cited above \cite{pogossian2022} 3 different definitions of the advance of Mercury's perihelion have been introduced. The first, which has been named APM-LRL, refers to the average rotation angle per century of the Laplace-Runge-Lenz vector with respect to the fixed reference direction. The second, which has been named extended APM in  \cite{pogossian2022}, refers to the angle between the reference direction and the position vector pointing to the extended perihelion at successive mean orbital periods of Mercury. The third, the geometric advance of the perihelion, called geometric APM, is calculated as the angle between the reference direction and the vector pointing from the nearest focus of the mean ellipse to the Sun to the point of the ellipse closest to this focus for each successive mean period. 
	
	The average advance per century of Mercury's perihelion is calculated over an initial adjustment period of 600 years which is gradually increased with a step of 5 years. This initial threshold of 600 years was chosen to damp the large amplitude oscillations of the extended and geometric APM  as described in reference \cite{pogossian2022}.
	
	\section{Results and Discussion}
	\label{sec:3}
	\
	Let us now analyze the hypothetical values of the orbital elements of the planet Vulcan reported by different authors. Le Verrier admitted a value for the semimajor axis equal to \textit{a}=0.1427 AU\cite{leverrier1860}. He considered a circular orbit. As for the position of the plane of motion of Vulcan with respect to the ecliptic, the values of the inclination and the longitude of the ascending node given by Le Verrier are  $ \textit i=12^\circ 10^{\prime}$ and  	$\Omega = 12^\circ 59^{\prime}$ respectively. 
	
	As indicated previously, the optimization of the orbital elements of the orbit of Vulcan consists in selecting values which allow having a maximal influence on the advance of the perihelion of Mercury, of course taking into account the results concerning the stability of such an orbit.  It is clear that the influence of Vulcan on the perihelion advance of Mercury will be maximal for the shortest distance between the two planets. Indeed, the APM is an increasing function of the decreasing distance between two stars.
	
	But, according to Evans et al.\cite{evans1999}  the stable orbits of the vulcanoids are only inside the orbit of Mercury between 0.09 and 0.20 AU, with a small gap at 0.15 AU and 0.18 AU corresponding to an unstable mean motion in resonance with Mercury and Venus.  Collisions are the main cause of such evolution of the vulcanoid belt. Large (> 1km) vulcanoids of several km are more likely to be found in the area \textit{a} < 0.18 AU. Evans and Tabachnik concluded that all long-term vulcanoids should orbit the Sun between 0.08 and 0.18 AU \cite{evans2002}, \cite{zhao2009}.  
	
	The observations of Stern and Durda \cite{stern2000} indicate that moving bodies are all the more stable on their orbit as their eccentricity is low. And the semimajor axis must be not far from 0.18 which is almost the outer edge of the dynamically stable region and this with highly circular orbits. These authors have also shown that only a few hundred objects with a radius greater than 1 km will survive if their average orbital eccentricity does not exceed $\sim10^{-3}$ 
	In the semimajor axis optimization procedure, the value of eccentricity was set to a very small value of $e = 10^{-3}$ according to the observations of  Stern and Durda \cite{stern2000}. Our preliminary calculation has shown that among the orbits, the interaction of Vulcan with Mercury and thus its influence on the perihelion advance of Mercury will be maximal if Vulcan is in the plane of Mercury's orbit, i.e. the same \textit{i} and $\Omega$ as Mercury's orbit. In the preliminary calculations, it was verified that the argument of the perihelion $\omega$ has practically no influence on the APM because of the very weak eccentricity of the orbit of Vulcan. The true initial anomaly of Vulcan has only a negligible influence on the three APMs at a time interval of one century, because of revolutions three times faster than those of Mercury around the sun.
	
	Moreover, the dependence of the APM on the longitude of the ascending node  of Vulcan is also very weak. In order to study the dependence of the APM on the semimajor axis \textit{a} and the eccentricity e of Vulcan, I have assigned to the longitude of the ascending node and to the argument of the perihelion of Vulcan respectively the same values as those of the orbit of Mercury. As for the true anomaly of Vulcan, I have arbitrarily attributed to it the value of $\nu=45.0^\circ$.  In these optimization calculations, the mass of Vulcan was considered equal to 1/17 of the mass of Mercury, as suggested by Le Verrier.  The table ~\ref{tab:Table 1} shows the dependence of Mercury's perihelia on the semimajor axis \textit{a} in the stability zone (0.08 AU < \textit{a} < 0.18 AU) defined by Evans. It can be seen from table ~\ref{tab:Table 1} that the optimal value of  the semimajor axis of the orbit of Vulcan, i.e., the value that gives the maximum value of the APM, is equal  to  \textit{a} = 0.18 AU.

	\begin{table*}
		\caption{This table shows how the different advances of the perihelion of Mercury depend on the value of the semimajor axis. In these optimization calculations, the mass of Vulcan has been considered equal to 1/17 of the mass of Mercury, as suggested by Le Verrier. The value of the eccentricity is considered to be 0.001.} 
		\label{tab:Table 1}
		\begin{center}
			\begin{tabular}{|p{2cm}||p{2.5cm}|p{2.5cm}|p{2.5cm}|} 
				\hline
				\multicolumn{4}{|c|}{\bf{Table 1}} \\
				\hline
				%				        &                                         \\ 
				\textbf{Semimajeur axis} \qquad \qquad {\textit{a} (AU)} & { \bf {APM-LRL}  {$(^{\prime\prime}/cy) $} } & {\bf {Extended APM}  {$(^{\prime\prime}/cy)$}} & {\bf{Geometric APM} {$(^{\prime\prime}/cy)$}}\\ 
				%		        &                                         \\ 
				%		\hline\hline
				\hline
				0.18 & 533.61 & 534.00 & 539.54\\ 
				\hline
				0.16 & 533.13 & 533.66 & 539.06\\
				\hline
				0.14 & 532.79 & 533.02 & 538.72\\
				\hline
				0.12 & 532.54 & 532.73 & 538.47\\
				\hline
				0.10 & 532.36 & 532.64 & 538.29\\ 
				\hline
				0.08 & 532.23 & 532.78 & 538.16\\  						
				\hline
			\end{tabular}
		\end{center}
	\end{table*}
	
	It has been mentioned that in the optimization procedure of the semimajor axis of the orbit of Vulcan, a very low eccentricity e=0.001 has been chosen according to the conclusions of Stern and Durda \cite{stern2000}.  But we will now proceed to the verification of this assumption. For that we chose for the value of semimajor axis its optimal value \textit{a}=0.18 AU and we have set the values of the longitude of the ascending node, the argument of perihelion as well as that of the true anomaly to the same values as in the optimization procedure of the semimajor axis. Table ~\ref{tab:Table 2} confirms that for fixed values of all other orbital elements, the highest APM is produced by low eccentricity orbits.  
	
	\begin{table*}
		\caption{This table shows how the different advances of the perihelion of Mercury depend on the value of the eccentricity. The mass of Vulcan has been considered equal to 1/17 of the mass of Mercury. I have used the optimized value of \textit{a}=0.18 AU of  semimajor axis. } 
		\label{tab:Table 2}
		\begin{center}
			\begin{tabular}{|p{2.5cm}||p{2.5cm}|p{2.5cm}|p{2.5cm}|} 
				\hline
				\multicolumn{4}{|c|}{\bf{Table 2}} \\
				\hline
				%				        &                                         \\ 
				\bf Eccentricity \qquad \qquad \textit{e}  & { \bf {APM-LRL}  {$(^{\prime\prime}/cy) $} } & {\bf {Extended APM}  {$(^{\prime\prime}/cy)$}} & {\bf{Geometric APM} {$(^{\prime\prime}/cy)$}}\\ 
				%		        &                                         \\ 
				%		\hline\hline
				\hline
				0.001 & 533.61 & 534.00 & 539.54\\ 
				\hline
				0.05 & 533.34 & 533.92 & 539.27\\
				\hline
				0.1 & 533.10 & 533.74 & 539.03\\
				\hline
				0.2 & 532.70 & 533.30 & 538.62\\	
				\hline
			\end{tabular}
		\end{center}
	\end{table*}
	
	By choosing the values \textit{a}=0.18 and e=0.001, one looks for the optimal positioning of the orbital plane of Vulcan with respect to the ecliptic. For which values of the orientation parameters of the orbit of Vulcan, i.e. the inclination angle and the longitude of the ascending node, the influence of Vulcan on the APM of the orbit of Mercury will be maximal.  The values recommended by Le Verrier for these quantities on the basis of Lescarbaults' observations (which turned out to be wrong) are  $ \textit i=12^\circ 10^{\prime}$ and  	$\Omega = 12^\circ 59^{\prime}$ respectively \cite{leverrier1860}, \cite{baum1997} page 156.  
	
	\begin{table*}
		\caption{This table shows how the different advances of the perihelion of Mercury depend on the value of the inclination angle of the orbit of Vulcan. The mass of Vulcan has been considered equal to 1/17 of the mass of Mercury.  I used the optimized values: \textit{a}=0.18 AU for the semimajor axis and e=0.001 for the eccentricity.} 
		\label{tab:Table 3}
		\begin{center}
			\begin{tabular}{|p{2.5cm}||p{2.5cm}|p{2.5cm}|p{2.5cm}|} 
				\hline
				\multicolumn{4}{|c|}{\bf{Table 3}} \\
				\hline
				%				        &                                         \\ 
				{\bf{ Inclination \textit{i} } {$(in\qquad degrees)$}}  & { \bf {APM-LRL}  {$(^{\prime\prime}/cy) $} } & {\bf {Extended APM}  {$(^{\prime\prime}/cy)$}} & {\bf{Geometric APM} {$(^{\prime\prime}/cy)$}}\\ 
				%		        &                                         \\ 
				%		\hline\hline
				\hline
				7.0138675-7 & 533.52 & 533.93 & 539.45\\ 
				\hline
				7.0138675-3.5 & 533.59 & 533.98 & 539.52\\
				\hline
				7.0138675 & 533.61 & 534.00 & 539.54\\
				\hline
				7.0138675+3.5 & 533.59 & 533.98 & 539.52\\	
				\hline
				7.0138675+7 & 533.52 & 533.92 & 539.45\\	
				\hline
			\end{tabular}
		\end{center}
	\end{table*}

	\begin{table*}
		\caption{This table shows how the different advances of the perihelion of Mercury depend on the longitude of the ascending node $\Omega$ of the orbit of Vulcan. The mass of Vulcan has been considered equal to 1/17 of the mass of Mercury.  I used the optimized values: \textit{a}=0.18 AU for the semimajor axis, e=0.001 for the eccentricity and \textit{i}=7.0138675$^\circ$ for the inclination of the orbit of Vulcan.
		} 
		\label{tab:Table 4}
		\begin{center}
			\begin{tabular}{|p{2.5cm}||p{2.5cm}|p{2.5cm}|p{2.5cm}|} 
				\hline
				\multicolumn{4}{|c|}{\bf{Table 4}} \\
				\hline
				%				        &                                         \\ 
				{\bf{Longitude of the ascending node} {$(in\qquad degrees)$}}  & { \bf {APM-LRL}  {$(^{\prime\prime}/cy) $} } & {\bf {Extended APM}  {$(^{\prime\prime}/cy)$}} & {\bf{Geometric APM} {$(^{\prime\prime}/cy)$}}\\ 
				%		        &                                         \\ 
				%		\hline\hline
				\hline
				48.1238724-30 & 533.59 & 533.97 & 539.51\\ 
				\hline
				48.1238724-15 & 533.60 & 533.62 & 539.53\\
				\hline
				48.1238724 & 533.61 & 534.00 & 539.54\\
				\hline
				48.1238724+15 & 533.60 & 534.31 & 539.53\\	
				\hline
				48.1238724+30 & 533.59 & 533.97 & 539.51\\	
				\hline
			\end{tabular}
		\end{center}
	\end{table*}
	
	Tables ~\ref{tab:Table 3} and ~\ref{tab:Table 4} show that the optimal inclination angle and the optimal longitude of the ascending node must be exactly equal to the corresponding values of Mercury's orbit, and that even a small deviation from this orientation of Vulcan's orbital plane will result in a decrease in Mercury's APM.
	
	In order to locate the elliptical trajectory within the orbital plane defined by \textit{i} and $\Omega$ it is necessary to give the position of perihelion with the perihelion argument $\omega$.  Since the eccentricity is considered negligible and therefore the trajectory of Vulcan is almost circular, the influence of the argument of perihelion $\omega$ on the APM of Mercury will be entirely neglected.  As already mentioned, its value was chosen arbitrarily equal to the corresponding value of the orbit of Mercury. The motion of Vulcan being faster than that of Mercury with a period of about one third of that of Mercury, there will be a very negligible dependence of the APM on the true anomaly of Vulcan and this is why the arbitrary value $\nu=45.0^\circ$ has been assigned to the true anomaly in all previous optimization calculations.
	
	As can be seen in its most favorable position of maximum influence, the APMs induced by Vulcan of mass 1/17th of that of Mercury as suggested by Le Verrier: APM-LRL $533.61^{\prime\prime}$/cy, extended APM $534.00^{\prime\prime}$/cy
	and geometric APM $539.54^{\prime\prime}$/cy are very far from the experimental value of $575.31^{\prime\prime}$/cy \cite{park2017}.  Therefore Vulcan with such a mass will not be able to explain the anomalous advance of Mercury's perihelion. Then what should be the mass of Vulcan for the induced APM to be consistent with its experimental value?  To answer this question, we must determine a mass of Vulcan such that it results in an anomalous advance of Mercury's perihelion equal to about $43^{\prime\prime}$/cy while keeping the optimal orbital parameters of Vulcan fixed.
	
	First, let us evaluate the APMs assuming the absence of Vulcan. The results obtained: APM-LRL $532.03^{\prime\prime}$/cy, extended APM $532.44^{\prime\prime}$/cy and geometric APM $537.96^{\prime\prime}$/cy are consistent with the values reported in the reference \cite{pogossian2022} with a slight difference due to the fact that for the calculation of APM per century the initial averaging interval was taken 600 years while in the latter reference it is 300 years.
	
	The dependence of the APM-LRL on the mass of Vulcan by fixing all other parameters follows a linear trend as shown in Fig 1. Thus, only for a Vulcan mass of about 1.6 times the mass of Mercury will we have the explanation of the anomalous APM-LRL in the Newtonian gravitational framework. For the extended APM, we have about 1.6 times the mass of Mercury, which gives the same magnitude. For the geometric APM, we have a slightly lower required mass of about 1.4 Mercury mass.  If we consider such a mass, Vulcan will be easily visible as Mercury since its angular size visible from the Earth will be of the order of magnitude 0.8 of that of Mercury.

	\begin{figure*}[!htb]
		\centering
		\resizebox{0.999\textwidth}{!}{\includegraphics[height=6.9 cm, width=7.2 cm,angle=0]{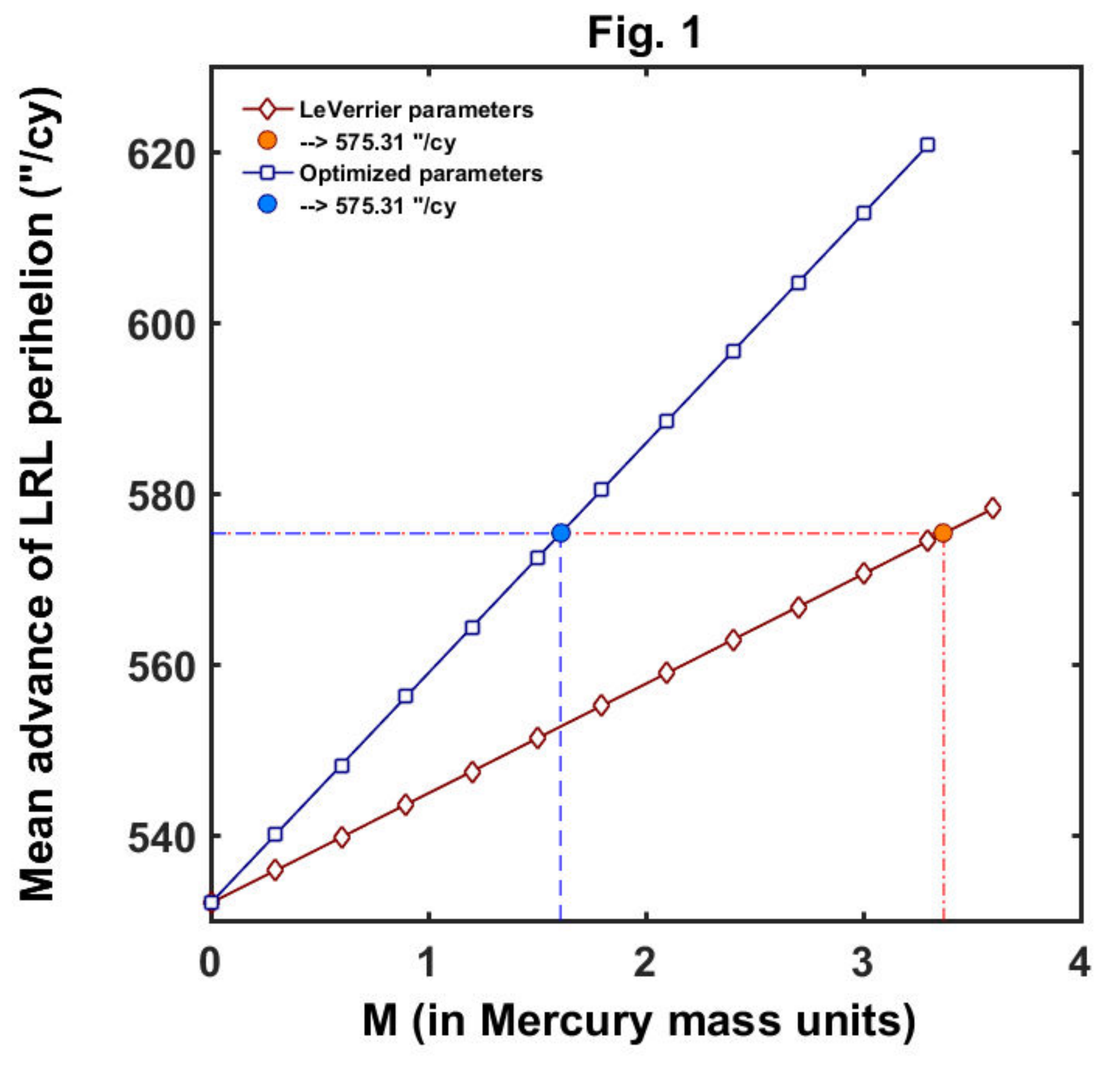}}
		\caption{\label{Fig. 1 :} The two curves with the square and diamond points correspond to the dependencies calculated using respectively the optimized orbital parameters and the parameters recommended by Le Verrier. The filled circles on the curve of each model correspond to the value of the perihelion advance of Mercury measured experimentally (575.31"/cy) whose level is marked by the horizontal dotted and dashed lines. The vertical dotted lines coming out of the filled circles of each model indicate the value of the mass of Vulcan corresponding to the experimental value of the APM-LRL.
		}
	\end{figure*}
	% Figure 1
	
	For comparison purposes, we have also plotted on Fig. 1 the dependence of the APM-LRL on the mass of Vulcan for another set of orbital parameters of the circular orbit of Vulcan corresponding to a value of \textit{a}= 0.1427 AU, $\textit i=12^\circ 10^{\prime}$ and $\Omega = 12^\circ 59^{\prime}$  proposed by Le Verrier. As for the value of $\nu$, it  has been fixed at the value previously specified in the optimization procedures of this work. As shown by the linear dependence of the perihelion advance defined by the Laplace Runge Lentz \cite{pogossian2022} vector rotation as a function of the mass of Vulcan in Fig. 1, only a Vulcan with 1.6 times the mass of Mercury can explain the anomalous perihelion advance of Mercury. The lower linear curve which is mainly based on the parameters recommended by Le Verrier will not be able to explain this anomalous advance of the perihelion of Mercury except to admit a mass for Vulcan more than about three times the mass of Mercury. In this case, Vulcan would be much more luminous than Mercury, all the more so as it is closer to the Sun and would therefore have been detected since antiquity.
	
	\section{Conclusion}
	Taking into account the recent progress in the experimental and theoretical exploration of the intra-mercurian zone supported by our calculations, we can conclude that the hypothetical Vulcan with a mass of 1/17 of Mercury and with realistic orbital parameters cannot explain the anomalous advance of Mercury's perihelion.  We optimized the orbital parameters of the hypothetical Vulcan in order to explain the anomalous advance of Mercury's perihelion. To account for this, Vulcan should have a mass of 1.6 times that of Mercury and would therefore be visible at almost the same solid angle as Mercury and could even be detectable since antiquity. 
	
	Modern observations also rule out the possibility of an intra-mercurian asteroid belt of fairly large total mass in the stability belt located in Mercury's orbit between 0.08 and 0.18 AU. The question of whether the planet Vulcan existed with a lower longevity than the bodies currently observed around the Sun remains open.

	%Bibliography
	\bibliographystyle{plain}

\end{document}